\documentclass[aps,pra,showpacs,a4paper]{revtex4}

\usepackage{amsmath,amsfonts,amssymb}
\usepackage{txfonts}
\usepackage{type1cm}

\begin{document}

\bibliographystyle{apsrev}

\title{A problem of existence of bound entangled states with non-positive partial transpose and the Hilbert's 17th problem
}

\author{Tohya Hiroshima}
\email{tohya@qci.jst.go.jp}
\affiliation{
Quantum Computation and Information Project, ERATO-SORST, Japan Science and Technology Agency,\\
Daini Hongo White Building 201, Hongo 5-28-3, Bunkyo-ku, Tokyo 113-0033, Japan.}

\begin{abstract}
It is found that the problem of existence of bound entangled states with non-positive partial transpose (NPPT) 
has the intriguing relation to the Hilbert's 17th problem.
More precisely, we compute the expectation value of the partially transposed Werner states by Schmidt rank-2 vectors 
for NPPT and 1-copy undistillable region.
It is the positive polynomial but shown not to be expressed as the sum of squares of polynomials.
A remedy for such pathological behavior as well as a similar but different approach to the problem is also mentioned.
\end{abstract}

\maketitle

\section{Introduction}

Quantum entanglement is at the heart of the quantum information science 
and is the name of the game of various types of quantum information processing \cite{NC,Alber,Hayashi}.
These fascinating tasks can be carried out with the highest performance 
by exploiting maximum entangled pure states. 
The pure entangled states are, however, very fragile and easily degraded to be mixed  
due to the unavoidable interaction with the environment, 
whereas, in a quantum communication setting, 
two or more distant parties are in principle not allowed to perform global operations 
on the shared entangled states; 
their ability is quite limited even though they can communicate each other.
As a remedy for such an unwanted decoherence effect of entanglement 
within the limits of distant party setting, 
the concept and protocols of distillation of entanglement, 
the procedure to extract pure maximally entangled states out of several copies of mixed states 
by using only local operations and classical communication (LOCC), 
have been introduced \cite{BBPSSW, BDSW,DEMPS,Gisin}.
At present, the distillability --- the possibility of distillation --- is in turn recognized 
as one of the most fundamental traits of entanglement 
\cite{PRHorodecki,Clarisse06a,RPMKHorodecki,PV}. 

The distillability \cite{comment_1} of bipartite states is provided by the following theorem.
A bipartite state $\rho $ acting on a composite Hilbert space 
$\mathcal{H}_{A}\otimes \mathcal{H}_{B}$ is distillable if and only if there exist 
a positive integer $N \in \mathbb{N}=\{1,2,\dots \}$ and 
a Schmidt rank-2 state vector $\left| \psi_{2}^{[N]}\right\rangle $ in 
$\mathcal{H}_{A}^{\otimes N}\otimes \mathcal{H}_{B}^{\otimes N}$ 
such that \cite{MPRHorodecki98}
\begin{equation} \label{eq:distillability}
\left\langle \psi_{2}^{[N]} \right| (\rho^{\otimes N})^{T_{B}}
\left| \psi_{2}^{[N]}\right\rangle = 
\left\langle \psi_{2}^{[N]} \right| 
(\rho ^{T_{B}})^{\otimes N} \left| \psi_{2}^{[N]} \right\rangle < 0. 
\end{equation}
Here, $T_{B}$ denotes the partial transpose with respect to system $B$ and 
the Schmidt rank-2 state vector is of the form, 
\begin{equation*}
\left| \psi_{2}^{[N]} \right\rangle =
\sum_{i=1,2}c_{i} 
\left| e_{i} \right\rangle _{A} \otimes \left| f_{i} \right\rangle _{B}
\end{equation*}
with $c_{i}\in \mathbb{R}$ and 
$\left| e_{i} \right\rangle _{A}$ ($\left| f_{i} \right\rangle _{B}$) ($i=1,2$) 
being orthonormal vectors in $\mathcal{H}_{A}^{\otimes N}$ ($\mathcal{H}_{B}^{\otimes N}$).

If Eq.~(\ref{eq:distillability}) is satisfied for some $N$, 
$\rho $ is called $N$-copy distillable.
Conversely, if it does not hold for up to some finite $N\in \mathbb{N}$, 
$\rho $ is called $N$-copy undistillable.
While $N$-copy distillable states are distillable by definition, 
the $N$-copy undistillability does not necessarily mean undistillability.
To show the undistillability, 
we need to prove the $N$-copy undistillability for all $N\in \mathbb{N}$.

Not all entangled states are distillable.
Undistillable and yet entangled states are called bound entangled (BE) states \cite{MPRHorodecki98}.
One notable example of BE states is 
a state $\rho $ whose partial transpose with respect to system $A(B)$ 
remains positive semi-definite; 
$\rho ^{T_{A(B)}} \geq 0$ \cite{MPRHorodecki98}.
Such a state is called a positive partial transpose (PPT) state.
Otherwise, it is called a non-PPT (NPPT) state.
It is clear from the above theorem that all PPT states are undistillable 
and that the distillable states are inevitably NPPT.
All distillable states known to date belong to the NPPT class.
So the natural question is: 
Is the NPPT condition also sufficient for distillability? or
is there an NPPT BE entangled state?
Since the NPPT condition is proven to be equivalent to distillability 
for two-qubit systems \cite{MPRHorodecki97} 
and for states on $\mathcal{H}_{A}\otimes \mathcal{H}_{B}$ 
with $\dim \mathcal{H}_{A}=2$ and $\dim \mathcal{H}_{B} \geq 3$ \cite{DCLB}, 
an NPPT BE entangled state, if it exists, must live in higher dimensions \cite{comment_2}.
This is a long-standing open problem and remains unanswered 
since the discovery of bound entanglement in 1998 \cite{Clarisse06a,RPMKHorodecki,PV}.
Although considerable efforts have been devoted in solving this problem so far 
\cite{DCLB,DSSTT,LBCKKSST,BCHHKLS,KLC,BR,Clarisse05,Clarisse06b,VD,Simon,CS,BE,PPHH}, 
the problem has been defying solution stubbornly.

Two early independent attempts \cite{DCLB,DSSTT} (see also \cite{LBCKKSST}) strongly suggest 
that NPPT Werner states lying at a narrow region adjacent to the PPT-NPPT borderline 
are undistillable; 
they have identified $N$-copy undistillable NPPT Werner states 
but the $N$-copy undistillable region is getting narrower and narrower and vanishes eventually 
as $N$ goes to infinity.
Although their findings may support the existence of NPPT BE Werner states, 
it is still inconclusive.
Guessing from the finite $N$ results may be quite misleading 
because there are states which are not $N$-copy distillable but $(N+1)$-copy distillable 
\cite{Watrous}.
The same is very much true for small $N$ numerical study \cite{VD}.
A Werner state \cite{Werner} in question acting on 
$\mathcal{H}_{A} \otimes \mathcal{H}_{B}$ with 
$\dim \mathcal{H}_{A}=\dim \mathcal{H}_{B}=d (\geq 3)$ takes the form,
\begin{equation*}
\rho_{W}(\alpha) = \frac{1}{d^{2}-d\alpha} (I-\alpha F), 
\end{equation*}
where 
$F=\sum_{i,j=1}^{d}\left| ij\right\rangle \left\langle ji\right| $
denotes the swap or flip operator and 
$-1\leq \alpha \leq 1$.
The partial transposition of $\rho_{W}(\alpha)$ reads
\begin{equation} \label{eq:Lambda}
\rho_{W}(\alpha)^{T_{B}} = \frac{1}{d^{2}-d\alpha} (I-d\alpha P_{+})=:
\frac{1}{d^{2}-d\alpha} \Lambda (\alpha), 
\end{equation}
where 
$P_{+}=\left| \Psi _{+}\right\rangle \left\langle \Psi _{+}\right| $ with 
$\left| \Psi _{+}\right\rangle =d^{-1/2}\sum_{i=1}^{d}\left| ii\right\rangle $.
For $-1\leq \alpha \leq 1/d$ $\rho_{W}(\alpha)$ is PPT and also separable, 
while for $1/d<\alpha \leq 1$ it is NPPT.
The NPPT region is further divided into two; 
for $1/2 < \alpha \leq 1$ $\rho_{W}(\alpha)$ is 1-copy distillable but for 
$1/d < \alpha \leq 1/2$ it is 1-copy undistillable, 
i.e., there does not exist a Schmidt rank-2 vector 
$\left| \psi_{2}^{[1]}\right\rangle \in \mathcal{H}_{A}\otimes \mathcal{H}_{B}$ 
such that 
$\left\langle \psi_{2}^{[1]} \right| \rho_{W}(\alpha)^{T_{B}}
\left| \psi_{2}^{[1]} \right\rangle <0$ 
in this range of parameter $\alpha$.
The conjecture is that for $1/d < \alpha \leq 1/2$ 
$\rho_{W}(\alpha)$ is also $N$-copy undistillable for all $N\in \mathbb{N}$, 
i.e., it is undistillable in an exact sense.

It is pointed out that any NPPT state is distillable 
if and only if any entangled Werner state is distillable \cite{MPHorodecki}.
The immediate consequence is as follows: 
if NPPT BE Werner states do not exist, 
then the NPPT condition is necessary and sufficient for distillability.
Thus, the solution would finalize the classification of (bipartite) quantum states 
from the viewpoint of entanglement whether the answer to the conjecture is positive or negative.
Furthermore, the existence of NPPT BE states has the following important implications \cite{SST}.
Namely, if they exist, then the distillable entanglement \cite{BDSW} 
--- one of the entanglement measures --- is not convex, and more surprisingly, 
the compelling yet counterintuitive phenomenon 
called superactivation \cite{Superactivation} is expected, 
that is, the tensor product of two BE states, 
one some PPT BE state and another the conjectured NPPT BE Werner state, 
turns out to be distillable.
It is furthermore shown in \cite{VW} that any NPPT state can be boosted to be 1-copy distillable 
by adding (tensor multiplying, to be exact) a PPT BE state 
with infinitesimal amount of entanglement.
This implies that, if the conjectured NPPT BE exists, 
then the superactivation is always triggered by an infinitesimally small amount of PPT BE.

In this paper, we point out that the problem of existence of NPPT-BE, which is still unanswered, is closely related to the Hilbert's 17th problem.
In Sec.~\ref{block_matirx}, we cast the original problem into a $2 \times 2$ block matrix form.
In Sec.~\ref{Hilbert}, we find the relation between the original problem and the Hilbert's 17th problem.
A similar but slightly different approach to the problem is also described.
Section \ref{Conclusions} is conclusions.

\section{$2 \times 2$ block matrix formulation} \label{block_matirx}

To begin with we write a Schmidt rank-2 state vector in 
$\mathcal{H}_{A}^{\otimes N}\otimes \mathcal{H}_{B}^{\otimes N}$ as 
$\left| \psi_{2}^{[N]}\right\rangle =\sum_{k=1,2}c_{k}\left| \psi_{k}\right\rangle $ 
with
\begin{equation*}
\left| \psi_{k} \right\rangle =
\sum_{i_{1},i_{2},\dots,i_{N}=1}^{d}
\sum_{j_{1},j_{2},\dots,j_{N}=1}^{d}
u_{i_{1}i_{2}\dots i_{N}}^{(k)}
v_{j_{1}j_{2}\dots j_{N}}^{(k)}
\left| i_{1}i_{2}\dots i_{N} \right\rangle_{A} \otimes 
        \left| j_{1}j_{2}\dots j_{N} \right\rangle_{B},
\end{equation*}
where $\sum_{k=1,2} c_{k}^{2}=1$, 
$\sum_{i_{1},i_{2},\dots,i_{N}=1}^{d} u_{i_{1}i_{2} \dots i_{N}}^{(k_{1})*}
u_{i_{1}i_{2} \dots i_{N}}^{(k_{2})}=\delta _{k_{1}k_{2}}$, 
and
$\sum_{j_{1},j_{2},\dots,j_{N}=1}^{d} v_{j_{1}j_{2} \dots j_{N}}^{(k_{1})*}
v_{j_{1}j_{2} \dots j_{N}}^{(k_{2})}=\delta _{k_{1}k_{2}}$ 
due to the orthonormal condition.
Our goal is to show 
$\left\langle \psi_{2}^{[N]} \right|
\Lambda (\alpha)^{\otimes N}
\left| \psi_{2}^{[N]}\right\rangle \geq 0$ 
for $1/d < \alpha \leq 1/2$ 
and for all $\left| \psi_{2}^{[N]} \right\rangle $, 
where $\Lambda (\alpha)=I-d\alpha P_{+} $ is defined in Eq.~(\ref{eq:Lambda}).
We write
\begin{equation} \label{eq:expectation}
\left\langle \psi_{2}^{[N]} \right| \Lambda (\alpha)^{\otimes N}
\left| \psi_{2}^{[N]} \right\rangle 
=\sum_{k_{1},k_{2}=1,2} c_{k_{1}}^{*}c_{k_{2}}
\sum_{i_{1},i_{2},\dots,i_{N}=1}^{d}
\sum_{j_{1},j_{2},\dots,j_{N}=1}^{d}
u_{i_{1}i_{2}\dots i_{N}}^{(k_{1})*}
\left( M_{(k_{1},k_{2})}^{[N]} \right) _{i_{1}i_{2}\dots i_{N},j_{1}j_{2} \dots j_{N}}
u_{j_{1}j_{2} \dots j_{N}}^{(k_{2})}
=:
\begin{pmatrix}
\mathbf{w}_{1} \\ 
\mathbf{w}_{2}
\end{pmatrix}
^{\dagger}
\begin{pmatrix}
M_{(1,1)}^{[N]} & M_{(1,2)}^{[N]} \\ 
M_{(2,1)}^{[N]} & M_{(2,2)}^{[N]}
\end{pmatrix}
\begin{pmatrix}
\mathbf{w}_{1} \\ 
\mathbf{w}_{2}
\end{pmatrix}
\end{equation}
with 
$\mathbf{w}_{k}=
(c_{k}u_{11\dots 1}^{(k)},c_{k}u_{11\dots 2}^{(k)},\dots,c_{k}u_{dd\dots d}^{(k)})^{T}$.
In Eq.~(\ref{eq:expectation}), matrix elements of 
$M_{(k_{1},k_{2})}^{[N]}$ are bilinear forms of 
$v_{i_{1}i_{2}\dots i_{N}}^{(k_{1})}$ and 
$v_{i_{1}i_{2}\dots i_{N}}^{(k_{2})}$ and 
$M_{(2,1)}^{[N]\dagger}=M_{(1,2)}^{[N]}$.

For $N=1$, it is easy to see 
$M_{(k_{1},k_{2})}^{[1]}=V_{k_{1}}^{[1]\dagger}\Lambda (\alpha)V_{k_{2}}^{[1]}$, 
where
\begin{equation} \label{eq:V_matrix_1}
V_{k}^{[1]}=
\begin{pmatrix}
\mathbf{v}^{(k)} & 0 & \dots  & 0 \\ 
0 & \mathbf{v}^{(k)} &  & \vdots  \\ 
\vdots  &  & \ddots  & \vdots  \\ 
0 & \dots  & \dots  & \mathbf{v}^{(k)}
\end{pmatrix}
\end{equation}
is a $d^{2}\times d$ matrix with 
$\mathbf{v}^{(k)}=(v_{1}^{(k)},v_{2}^{(k)},\dots,v_{d}^{(k)})^{T}$, 
by which one finds
\begin{align} \label{eq:m1_tensor}
&\left( M_{(k_{1},k_{2})}^{[1]} \right)_{i_{1}i_{2} \dots i_{N},j_{1}j_{2} \dots j_{N}}^{\otimes N} \nonumber \\
&=\sum_{l_{1},l_{2},\dots,l_{N}=1}^{d}
 \sum_{m_{1},m_{2},\dots,m_{N}=1}^{d}
(V_{k_{1}}^{[1] \dagger})_{i_{1}l_{1}} (V_{k_{1}}^{[1]\dagger})_{i_{2}l_{2}}
\dots (V_{k_{1}}^{[1]\dagger})_{i_{N}l_{N}}
\Lambda (\alpha)_{l_{1}l_{2} \dots l_{N},m_{1}m_{2} \dots m_{N}}^{\otimes N}
\left( V_{k_{2}}^{[1]}\right)_{m_{1}j_{1}} \left( V_{k_{2}}^{[1]} \right)_{m_{2}j_{2}}
\dots \left( V_{k_{2}}^{[1]} \right)_{m_{N}j_{N}}.
\end{align}
This is the coefficient of 
$u_{i_{1}}^{(k_{1})*}u_{i_{2}}^{(k_{1})*} \dots u_{i_{N}}^{(k_{1})*}
u_{j_{1}}^{(k_{2})}u_{j_{2}}^{(k_{2})} \dots u_{j_{N}}^{(k_{2})}$ 
in 
$\left\langle \phi_{k_{1}} \right| \Lambda (\alpha)^{\otimes N}
\left| \phi_{k_{2}} \right\rangle $ 
with 
\begin{equation*}
\left| \phi _{k}\right\rangle =
\left( \sum_{i,j=1}^{d} u_{i}^{(k)} \left| i \right\rangle_{A} \otimes v_{j}^{(k)}
\left| j \right\rangle _{B} \right)^{\otimes N}.
\end{equation*}
Hence, if we replace 
$v_{i_{1}}^{(k)}v_{i_{2}}^{(k)} \dots v_{i_{N}}^{(k)}$ 
to 
$v_{i_{1}i_{2} \dots i_{N}}^{(k)}$ 
in 
$V_{k}^{[1] \otimes N}$ 
and read 
$u_{i_{1}}^{(k)}u_{i_{2}}^{(k)} \dots u_{i_{N}}^{(k)}$ 
as 
$u_{i_{1}i_{2} \dots i_{N}}^{(k)}$, Eq.~(\ref{eq:m1_tensor}) gives the coefficient of 
$u_{i_{1}i_{2} \dots i_{N}}^{(k_{1})*} u_{j_{1j}i_{2} \dots j_{N}}^{(k_{2})}$ 
in 
$\left\langle \psi_{k_{1}} \right| \Lambda (\alpha)^{\otimes N}
\left| \psi_{k_{2}} \right\rangle $.
Consequently, we can write
\begin{equation*}
M_{(k_{1},k_{2})}^{[N]} = V_{k_{1}}^{[N]\dagger}
\Lambda (\alpha)^{\otimes N} V_{k_{2}}^{[N]}, 
\end{equation*}
where $V_{k}^{[N]}$ is a $d^{2N} \times d^{N}$ matrix obtained from $V_{k}^{[1]\otimes N}$ 
by the above prescription.
By this construction of the linear map, 
$\mathbf{v}^{(k)}\in \mathbb{C}^{d^{n}}\mapsto V_{k}^{[N]}$, 
we can write 
\begin{equation} \label{eq:V_matrix_N}
V_{k}^{[N]} = \sum_{i_{1},i_{2},\dots,i_{N}=1}^{d}
\zeta^{(k)}(i_{1},i_{2},\dots,i_{N})
V_{i_{1}} \otimes V_{i_{2}} \otimes \dots \otimes V_{i_{N}}
\end{equation}
with 
$\sum_{i_{1},i_{2},\dots,i_{N}=1}^{d}
\left| \zeta^{(k)}(i_{1},i_{2},\dots,i_{N}) \right|^{2} = 1$, 
according to the expansion of the normalized vector $\mathbf{v}^{(k)}$: 
\begin{equation*}
\mathbf{v}^{(k)} = \sum_{i_{1},i_{2},\dots,i_{N}=1}^{d}
\zeta^{(k)}(i_{1},i_{2},\dots,i_{N})
\mathbf{v}_{i_{1}} \otimes \mathbf{v} _{i_{2}} \otimes \dots \otimes \mathbf{v}_{i_{N}} ,
\end{equation*}
where 
$\left\{ \mathbf{v}_{1},\mathbf{v}_{2},\dots,\mathbf{v}_{d}\right\} $ 
is a set of orthonormal basis of the $d$-dimensional complex vector space $\mathbb{C}^{d}$.
In Eq.~(\ref{eq:V_matrix_N}), $V_{i}$ is given by Eq.~(\ref{eq:V_matrix_1}) with 
$\mathbf{v}^{(k)} \rightarrow \mathbf{v}_{i}$.

It is convenient to expand a (normalized) vector $\left| x_{k}^{[N]} \right\rangle $ 
in $\mathbb{C}^{d^{N}}$ on which the operator $M_{(k_{1},k_{2})}^{[N]}$ acts as follows.
\begin{equation*}
\left| x_{k}^{[N]} \right\rangle =
\sum_{i_{1},i_{2},\dots,i_{N}=1}^{d}
\eta^{(k)}(i_{1},i_{2},\dots,i_{N}) \left| x_{i_{1}} \right\rangle \otimes
\left| x_{i_{2}} \right\rangle \otimes \dots \otimes 
\left| x_{i_{N}} \right\rangle .
\end{equation*}
Here, 
$\sum_{i_{1},i_{2},\dots,i_{N}=1}^{d}
\left| \eta^{(k)}(i_{1},i_{2},\dots,i_{N}) \right|^{2}=1$ 
and 
$\left| x_{i} \right\rangle = \mathbf{v}_{i}^{*}$ 
($i=1,2,\dots,d$), 
which also constitute a set of orthonormal basis of $\mathbb{C}^{d}$.
This choice of $\left| x_{i} \right\rangle $ is 
due to the observation that $\mathbf{v}_{i}^{*}$ is the eigenvector of 
$V_{i}^{\dagger} \Lambda (\alpha) V_{i}$ with the eigenvalue $1-\alpha$.
This fact is easily verified by direct computations.
Writing the vector $V_{j} \left| x_{i} \right\rangle $ explicitly as
\begin{equation*}
V_{j}\mathbf{v}_{i}^{*}=
\begin{pmatrix}
\mathbf{v}_{j} & 0 & \dots  & 0 \\ 
0 & \mathbf{v}_{j} &  & \vdots  \\ 
\vdots  &  & \ddots  & \vdots  \\ 
0 & \dots  & \dots  & \mathbf{v}_{j}
\end{pmatrix}
\begin{pmatrix}
(v_{i})_{1}^{*} \\ 
(v_{i})_{2}^{*} \\ 
\vdots  \\ 
(v_{i})_{d}^{*}
\end{pmatrix}
=
\begin{pmatrix}
(v_{i})_{1}^{*}\mathbf{v}_{j} \\ 
(v_{i})_{2}^{*}\mathbf{v}_{j} \\ 
\vdots  \\ 
(v_{i})_{d}^{*}\mathbf{v}_{j}
\end{pmatrix}, 
\end{equation*}
the inner product between this and the vector $\left| pp \right\rangle $ 
is readily computed as 
$\left\langle pp \right| V_{j}\mathbf{v}_{i}^{*} = (v_{i})_{p}^{*}(v_{j})_{p}$ 
for $p=1,2,\dots,d$, 
by which one can show that 
\begin{equation*}
\left\langle x_{i}\right| V_{j}^{\dagger} \Lambda (\alpha)
V_{l} \left| x_{m} \right\rangle  
=\mathbf{(v}_{i}^{*}\mathbf{)}^{\dagger}V_{j}^{\dagger} \Lambda (\alpha)
V_{l} \mathbf{v}_{m}^{*}
=\mathbf{v}_{i}^{T} V_{j}^{\dagger} V_{l} \mathbf{v}_{m}^{*}-\alpha
\sum_{p=1}^{d} (v_{i})_{p} (v_{j})_{p}^{*}
\sum_{q=1}^{d} (v_{l})_{q} (v_{m})_{q}^{*}
=\delta _{im} \delta _{jl} - \alpha \delta _{ij} \delta _{lm} .
\end{equation*}

In the following, we investigate the positivity of 
$\left\langle \psi_{2}^{[N]}\right| \Lambda (\alpha)^{\otimes N}\left|
\psi_{2}^{[N]}\right\rangle $, i.e.,
\begin{equation}
M^{[N]}(\alpha):=
\begin{pmatrix}
M_{(1,1)}^{[N]} & M_{(1,2)}^{[N]} \\ 
M_{(2,1)}^{[N]} & M_{(2,2)}^{[N]}
\end{pmatrix}
\geq 0, 
\end{equation}
or equivalently
\begin{equation} \label{eq:f_d}
f_{d}^{[N]}(\alpha ;\mathbf{v}^{(1)},\mathbf{v}^{(2)},\mathbf{w}^{(1)},\mathbf{w}^{(2)})
:=
\begin{pmatrix}
\mathbf{w}^{(1)} \\ 
\mathbf{w}^{(2)}
\end{pmatrix}
^{\dagger }
\begin{pmatrix}
M_{(1,1)}^{[N]} & M_{(1,2)}^{[N]} \\ 
M_{(2,1)}^{[N]} & M_{(2,2)}^{[N]}
\end{pmatrix}
\begin{pmatrix}
\mathbf{w}^{(1)} \\ 
\mathbf{w}^{(2)}
\end{pmatrix}
\geq 0
\end{equation}
for all $\mathbf{v}^{(1)}$, $\mathbf{v}^{(2)}$, $\mathbf{w}^{(1)}$, and $\mathbf{w}^{(2)}$
when $1/d \leq \alpha \leq 1/2$.

\section{Search for positive semi-definiteness of $M^{[N]}(\alpha)$ and the Hilbert's 17th problem} \label{Hilbert}

If the polynomial $f_{d}^{[N]}$ on $\mathbf{v}^{(1)}$, $\mathbf{v}^{(2)}$, $\mathbf{w}^{(1)}$, and $\mathbf{w}^{(2)}$ 
defined by Eq.~(\ref{eq:f_d}) were written as the sum of squares of the absolute values of polynomials, it would be positive semi-definite.
However, this is not the case even for $N=1$ in which $f_{d}^{[N=1]}$ is known to be positive semi-definite when $1/d \leq \alpha \leq 1/2$ \cite{DCLB,DSSTT}.
To see this we assume that all variables involved are real and set $d=3$.
Furthermore, we set
$x_{3}^{(1)}=x_{3}^{(2)}=y_{3}^{(1)}=y_{3}^{(2)}=z\in \mathbb{R}$
for simplicity.
The polynomial $f_{d=3}^{[N=1]}(\alpha =1/2)$ is now computed as
\begin{align} \label{eq:f}
&f_{d=3}^{[N=1]}(\alpha =1/2;\mathbf{v}^{(1)},\mathbf{v}^{(2)},\mathbf{w}^{(1)},\mathbf{w}^{(2)}) 
=f(v_{1}^{(1)},v_{2}^{(1)},v_{1}^{(2)}v_{2}^{(2)},w_{1}^{(1)},w_{2}^{(1)},w_{1}^{(2)}w_{2}^{(2)},z) \nonumber \\
&=2z^{4}+z^{2}v_{1}^{(1)2}+z^{2}v_{2}^{(1)2}+2z^{2}v_{1}^{(1)}v_{1}^{(2)}+z^{2}v_{1}^{(2)2}+2z^{2}v_{2}^{(1)}v_{2}^{(2)}
+z^{2}v_{2}^{(2)2}-2z^{2}v_{1}^{(1)}w_{1}^{(1)}+z^{2}w_{1}^{(1)2}+\frac{1}{2}v_{1}^{(1)2}w_{1}^{(1)2}+v_{2}^{(1)2}w_{1}^{(1)2} \nonumber \\
&-2z^{2}v_{2}^{(1)}w_{2}^{(1)}-v_{1}^{(1)}v_{2}^{(1)}w_{1}^{(1)}w_{2}^{(1)}+z^{2}w_{2}^{(1)2}+v_{1}^{(1)2}w_{2}^{(1)2}
+\frac{1}{2}v_{2}^{(1)2}w_{2}^{(1)2} -2z^{2}v_{1}^{(2)}w_{1}^{(2)}+2z^{2}w_{1}^{(1)}w_{1}^{(2)}+v_{1}^{(1)}v_{1}^{(2)}w_{1}^{(1)}w_{1}^{(2)}\nonumber \\
&+2v_{2}^{(1)}v_{2}^{(2)}w_{1}^{(1)}w_{1}^{(2)}-v_{2}^{(1)}v_{1}^{(2)}w_{2}^{(1)}w_{1}^{(2)}+z^{2}w_{1}^{(2)2}
+\frac{1}{2}v_{1}^{(2)2}w_{1}^{(2)2}+v_{2}^{(2)2}w_{1}^{(2)2}-2z^{2}v_{2}^{(2)}w_{2}^{(2)}-v_{1}^{(1)}v_{2}^{(2)}w_{1}^{(1)}w_{2}^{(2)}
+2z^{2}w_{2}^{(1)}w_{2}^{(2)}
\nonumber \\
&+2v_{1}^{(1)}v_{1}^{(2)}w_{2}^{(1)}w_{2}^{(2)}+v_{2}^{(1)}v_{2}^{(2)}w_{2}^{(1)}w_{2}^{(2)}-v_{1}^{(2)}v_{2}^{(2)}w_{1}^{(2)}w_{2}^{(2)}
+z^{2}w_{2}^{(2)2}+v_{1}^{(2)2}w_{2}^{(2)2}+\frac{1}{2}v_{2}^{(2)2}w_{2}^{(2)2}
\end{align}
This is the homogeneous real polynomial of $n_{var}=9$ real variables with even degree $2d_{f}=4$.

We follow the relaxation method \cite{PS} to denote the column vector whose entries are all the monomials in 
$v_1^{(1)}$, $v_2^{(1)}$, $\dots$, $w_2^{(2)}$, and $z$ of degree at most $d_{f}=2$ as $X$.
Furthermore, let $\mathcal{L}_{f}$ denote the set of all real symmetric $n\times n$ matrices $M(\alpha =1/2)$ such that 
$f=X^{T}M(\alpha =1/2)X$,
where $n$ is given by 
\[
n=
\begin{pmatrix}
n_{var}+d_{f} \\ 
d_{f}
\end{pmatrix}
=
\begin{pmatrix}
11 \\ 
2
\end{pmatrix}
=55.
\]
If there exists a positive semi-definite matrix $M(\alpha =1/2)\geq 0$ within ths set $\mathcal{L}_{f}$,
$f$ is written as the sum of squares of polynomials (SOS form).
Our task is to find such a positive semi-definite matrix.
Note that the monomial entries of $X$ with degree other than two can be omitted from the start 
because the squares of such components do not contribute to the {\it homogeneous\/} polynomial $f$ of degree four;
the corresponding diagonal elements of $M(\alpha =1/2)$ are all zero.
Thus, the vector $X$ is reduced to the following.
\begin{align*}
X &=\left\{
z^{2},zv_{1}^{(1)},zv_{2}^{(1)},zv_{1}^{(2)},zv_{2}^{(2)},zw_{1}^{(1)},zw_{2}^{(1)},zw_{1}^{(2)},zw_{2}^{(2)},\right. 
\\
&\left.
v_{1}^{(1)}w_{1}^{(1)},v_{1}^{(1)}w_{2}^{(1)},v_{2}^{(1)}w_{1}^{(1)},v_{2}^{(1)}w_{2}^{(1)},v_{1}^{(2)}w_{1}^{(2)},v_{1}^{(2)}w_{2}^{(2)},v_{2}^{(2)}w_{1}^{(2)},v_{2}^{(2)}w_{2}^{(2)}\right\},
\end{align*}
The reduced set $\mathcal{L}_{f}$ consists of all $17 \times 17$ matrices with $c_{i,j} \in \mathbb{R}$ as follows.
\[
M\left( \alpha =\frac{1}{2}\right) =\frac{1}{2}
\begin{pmatrix}
M_{A} & M_{C} \\ 
M_{C}^{T} & M_{B}
\end{pmatrix},
\]
where
\[
M_{A}=
\begin{pmatrix}
M_{A}^{1,1} & M_{A}^{1,2} \\ 
M_{A}^{1,2T} & M_{A}^{2,2}
\end{pmatrix}
\]
with
\[
M_{A}^{1,1}=
\begin{pmatrix}
4 & 0 & 0 & 0 & 0 \\ 
0 & 2 & 0 & 2 & 0 \\ 
0 & 0 & 2 & 0 & 2 \\ 
0 & 2 & 0 & 2 & 0 \\ 
0 & 0 & 2 & 0 & 2
\end{pmatrix},
\quad
M_{A}^{2,2}=
\begin{pmatrix}
2 & 0 & 2 & 0 \\ 
0 & 2 & 0 & 2 \\ 
2 & 0 & 2 & 0 \\ 
0 & 2 & 0 & 2
\end{pmatrix},
\]
and
\[
M_{A}^{1,2}=
\begin{pmatrix}
0 & 0 & 0 & 0 \\ 
-2-c_{1,10} & -c_{1,11} & 0 & 0 \\ 
-c_{1,12} & -2-c_{1,13} & 0 & 0 \\ 
0 & 0 & -2-c_{1,14} & -c_{1,15} \\ 
0 & 0 & -c_{1,16} & -2-c_{1,17}
\end{pmatrix},
\]

\[
M_{B}=
\begin{pmatrix}
M_{B}^{1,1} & M_{B}^{1,2} \\ 
M_{B}^{1,2T} & M_{B}^{2,2}
\end{pmatrix}
\]
with
\[
M_{B}^{1,1}=
\begin{pmatrix}
1 & 0 & 0 & c_{10,13} \\ 
0 & 2 & -1-c_{10,13} & 0 \\ 
0 & -1-c_{10,13} & 2 & 0 \\ 
c_{10,13} & 0 & 0 & 1
\end{pmatrix},
\]
\[
M_{B}^{2,2}=
\begin{pmatrix}
1 & 0 & 0 & c_{14,17} \\ 
0 & 2 & -1-c_{14,17} & 0 \\ 
0 & -1-c_{14,17} & 2 & 0 \\ 
c_{14,17} & 0 & 0 & 1
\end{pmatrix},
\]
and 
\[
M_{B}^{1,2}=
\begin{pmatrix}
1 & 0 & 0 & -1 \\ 
0 & 2 & 0 & 0 \\ 
0 & 0 & 2 & 0 \\ 
-1 & 0 & 0 & 1
\end{pmatrix},
\]
and 
\[
M_{C}=
\begin{pmatrix}
c_{1,10} & c_{1,11} & c_{1,12} & c_{1,13} & c_{1,14} & c_{1,15} & c_{1,16} & 
c_{1,17} \\ 
0 & 0 & c_{2,12} & c_{2,13} & 0 & 0 & 0 & 0 \\ 
-c_{2,12} & -c_{2,13} & 0 & 0 & 0 & 0 & 0 & 0 \\ 
0 & 0 & 0 & 0 & 0 & 0 & c_{4,16} & c_{4,17} \\ 
0 & 0 & 0 & 0 & -c_{4,16} & -c_{4,17} & 0 & 0 \\ 
0 & c_{6,11} & 0 & c_{6,13} & 0 & 0 & 0 & 0 \\ 
-c_{6,11} & 0 & -c_{6,13} & 0 & 0 & 0 & 0 & 0 \\ 
0 & 0 & 0 & 0 & 0 & c_{8,15} & 0 & c_{8,17} \\ 
0 & 0 & 0 & 0 & -c_{8,15} & 0 & -c_{8,17} & 0
\end{pmatrix}.
\]
One can check the identity 
$f=X^{T}M(\alpha =1/2)X$
by direct computations.
To investigate the positivity of $M(\alpha =1/2)$, we make following replacements, 
$c_{1,10}\rightarrow c_{1},c_{1,11}\rightarrow c_{2},\ldots ,c_{14,17}\rightarrow c_{18}$,
for simplicity.
Here we denote the principal submatrix (PSM) that lies in the rows and columns of $M(\alpha =1/2)$ indexed by ${i,j,k}$ as $M_{\{i,j,k\}}$.
We recall here that the necessary and sufficient condition of $M(\alpha =1/2) \geq 0$ is that 
all the PSM's of $M(\alpha =1/2)$ are positive semi-definite \cite{HJ}.
The PSM $M_{\{2,6,8\}}$ takes the form
\[
M_{\{2,6,8\}}=
\begin{pmatrix}
2 & -2-c_{1} & 0 \\ 
-2-c_{1} & 2 & 2 \\ 
0 & 2 & 2
\end{pmatrix}.
\]
Requiring that all the eigenvalues of $M_{\{2,6,8\}}$ be non-negative, we have $c_{1}=-2$.
Similar procedures determine remaining variables from $c_{2}$ to $c_{18}$.
The results are listed in Table~\ref{tab:c}.

\begin{table}[h] \label{tab:c}
\caption{Values of $c _i$}
\begin{center}
\begin{tabular}{ll|ll} \hline
{PSM} & {variable} & {PSM} & {variable}\\ \hline
$M_{\{2,6,8\}}$   & $c_{1}=-2$ & $M_{\{3,11,15\}}$  & $c_{10}=0$  \\
$M_{\{2,7,9\}}$   & $c_{2}=0$  & $M_{\{4,12,16\}}$  & $c_{11}=0$  \\
$M_{\{3,6,8\}}$   & $c_{3}=0$  & $M_{\{5,11,15\}}$  & $c_{12}=0$  \\
$M_{\{3,7,9\}}$   & $c_{4}=-2$ & $M_{\{6,11,15\}}$  & $c_{13}=0$  \\
$M_{\{4,6,8\}}$   & $c_{5}=-2$ & $M_{\{7,12,16\}}$  & $c_{14}=0$  \\
$M_{\{4,7,9\}}$   & $c_{6}=0$  & $M_{\{8,11,15\}}$  & $c_{15}=0$  \\
$M_{\{5,6,8\}}$   & $c_{7}=0$  & $M_{\{9,12,16\}}$  & $c_{16}=0$  \\
$M_{\{5,7,9\}}$   & $c_{8}=-2$ & $M_{\{11,12,16\}}$ & $c_{17}=-1$ \\
$M_{\{1,12,16\}}$ & $c_{9}=0$  & $M_{\{12,15,16\}}$ & $c_{18}=-1$ \\
\hline
\end{tabular}
\end{center}
\end{table}

Using these values of $c_i$ ($i=1,2,\dots, 18$), the minimal eigebvalue of $M(\alpha =1/2)$ is found to be 
$1-\sqrt{5}<0$, which means that positive semi-definite polynomial $f$ does not take the SOS form.

For $\alpha =1/d=1/3$, we also define the matrix $M(\alpha =1/3)$ such that 
$f_{d=3}^{[N=1]}(\alpha =1/3)=X^{T}M(\alpha =1/3)X$.
Similar procedures described above lead to the following form for $M(\alpha =1/3)$;
\[
M\left( \alpha =\frac{1}{3}\right) =\frac{1}{3}
\begin{pmatrix}
M_{A} & M_{C} \\ 
M_{C}^{T} & M_{B}
\end{pmatrix},
\]
where
\[
M_{A}=
\begin{pmatrix}
8 & 0 & 0 & 0 & 0 & 0 & 0 & 0 & 0 \\ 
0 & 3 & 0 & 3 & 0 & 0 & 0 & 0 & 0 \\ 
0 & 0 & 3 & 0 & 3 & 0 & 0 & 0 & 0 \\ 
0 & 3 & 0 & 3 & 0 & 0 & 0 & 0 & 0 \\ 
0 & 0 & 3 & 0 & 3 & 0 & 0 & 0 & 0 \\ 
0 & 0 & 0 & 0 & 0 & 3 & 0 & 3 & 0 \\ 
0 & 0 & 0 & 0 & 0 & 0 & 3 & 0 & 3 \\ 
0 & 0 & 0 & 0 & 0 & 3 & 0 & 3 & 0 \\ 
0 & 0 & 0 & 0 & 0 & 0 & 3 & 0 & 3
\end{pmatrix},
\]
\[
M_{B}=
\begin{pmatrix}
2 & 0 & 0 & -1 & 2 & 0 & 0 & -1 \\ 
0 & 3 & 0 & 0 & 0 & 3 & 0 & 0 \\ 
0 & 0 & 3 & 0 & 0 & 0 & 3 & 0 \\ 
-1 & 0 & 0 & 2 & -1 & 0 & 0 & 2 \\ 
2 & 0 & 0 & -1 & 2 & 0 & 0 & -1 \\ 
0 & 3 & 0 & 0 & 0 & 3 & 0 & 0 \\ 
0 & 0 & 3 & 0 & 0 & 0 & 3 & 0 \\ 
-1 & 0 & 0 & 2 & -1 & 0 & 0 & 2
\end{pmatrix},
\]
and
\[
M_{C}=
\begin{pmatrix}
-2 & 0 & 0 & -2 & -2 & 0 & 0 & -2 \\ 
0 & 0 & 0 & 0 & 0 & 0 & 0 & 0 \\ 
0 & 0 & 0 & 0 & 0 & 0 & 0 & 0 \\ 
0 & 0 & 0 & 0 & 0 & 0 & 0 & 0 \\ 
0 & 0 & 0 & 0 & 0 & 0 & 0 & 0 \\ 
0 & 0 & 0 & 0 & 0 & 0 & 0 & 0 \\ 
0 & 0 & 0 & 0 & 0 & 0 & 0 & 0 \\ 
0 & 0 & 0 & 0 & 0 & 0 & 0 & 0 \\ 
0 & 0 & 0 & 0 & 0 & 0 & 0 & 0
\end{pmatrix}.
\]
The minimum eigenvalue of $M(\alpha =1/3)$ is found to be zero.
Since $f_d^{[N=1]}(\alpha)$ is linear with respect to $\alpha $, the maximum of $\lambda _{min}(M(\alpha))$, 
the minimum eigenvalue of $M(\alpha)$, with respect to $c_i$ ($i=1,2,\dots,18$) is negative for $1/d < \alpha \leq 1/2$, 
the positive semi-definite polynomial $f_{d=3}^{[N=1]}(\alpha )$ cannot be expressed as an SOS form for $1/3 \leq \alpha \leq 1/2$.

The problem of existence of such positive semi-definite polynomials that cannot be written as an SOS form 
is related to the Hilbert's 17th problem that dates back to 1900.
The Hilbert's 17th problem was solved affirmatively by Artin in 1927.

{\it Theorem (Artin)} --- 
Every polynomial $p \in \mathbb{R}[\mathbf{x}]$ that is nonnegative on $\mathbb{R}^n$ is a sum of squares of rational functions, i.e., 
$p=\sum_{j}(a_{j}/b_{j})^{2}$ for some $a_{j},b_{j}\in \mathbb{R}[\mathbf{x}]$.

Here, one may ask a question.
Is it possible to write every positive semi-definite polynomial as a sum of squares of polynomials?
Unfortunately this is not true.
A series of counterexamples to this question was firstly found in 1965 by Motzkin.
One of the special cases of Motzkin's polynomials takes the form
\[
p_{M}=x_{1}^{2}x_{2}^{2}(x_{1}^{2}+x_{2}^{2}-3)+1,
\]
which is nonnegative but cannot be expressed as an SOS form.
Since then many such polynomials are found.
Our polynomial $f$ [Eq.~(\ref{eq:f})] consists one of them.
See \cite{Rez00} for an excellent survey on the Hilbert's 17th problem and the historical remarks including Motzkin's polynomials.
To show the positivity of $f$, we have to construct a set of rational functions satisfying Artin's theorem.
This is actually quite a formidable task.
However, Reznik proved the following recently \cite{Rez95}.

{\it Theorem (Reznik)} --- 
Let $p\in \mathbb{R}[\mathbf{x}]$ be a positive homogeneous polynomial on $\mathbb{R}^n$.
If $p>0$ on $\mathbb{R}\setminus \{0\}$, then there exists $r \in \mathbb{N}$ for which the polynomial 
$\left( \sum_{i=1}^{n}x_{i}^{2}\right) ^{r}p$ 
is a sum of squares.

This theorem might be helpful to prove $f_d^{[N]} \geq 0$.
However, small $r$ trial does not seem to work well even for $N=1$ and real variables case.
What is the minimum $r=r(N,d)$ to show $f_d^{[N]} \geq 0$?
Is it impossible for general $N$?
These questions remains to be answered.

So far we have utilized Eq.~(\ref{eq:f_d}) to investigate the positivity of $M^{[N]}(\alpha)$.
This is not an only option.
The inequality [Eq.~(\ref{eq:f_d})] is equivalent to the following \cite{HJ}.
(A) $M_{(1,1)}^{[N]}>0$ and 
(B) $\left\langle x_{1}^{[N]}\right| M_{(1,1)}^{[N]} \left| x_{1}^{[N]} \right\rangle 
\left\langle x_{2}^{[N]} \right| M_{(2,2)}^{[N]} \left| x_{2}^{[N]} \right\rangle \geq 
\left| \left\langle x_{1}^{[N]} \right|
M_{(1,2)}^{[N]} \left| x_{2}^{[N]} \right\rangle \right|^{2}$ 
for all 
$\left| x_{1}^{[N]} \right\rangle , \left| x_{2}^{[N]} \right\rangle 
\in \mathbb{C} ^{d^{N}}$.

The first inequality is proven as follows.
We decompose $\Lambda (\alpha )$ into $\Lambda (\alpha )=(1-\alpha )I+\alpha Z$ so that
\begin{align} \label{eq:sum}
&\left\langle x_{1}^{[N]}\right| V_{1}^{[N]\dagger }\Lambda (\alpha)^{\otimes N}V_{1}^{[N]}\left| x_{1}^{[N]}\right\rangle \nonumber \\
&=\sum_{n=0}^{N}(1-\alpha )^{n}\alpha ^{N-n}
\left( \left\langle x_{1}^{[N]}\right| V_{1}^{[N]\dagger }I^{\otimes n}\otimes Z^{\otimes (N-n)}V_{1}^{[N]}\left| x_{1}^{[N]}\right\rangle
+\ldots \left\langle x_{1}^{[N]}\right| V_{1}^{[N]\dagger }Z^{\otimes (N-n)}\otimes I^{\otimes n}V_{1}^{[N]}\left| x_{1}^{[N]}\right\rangle \right).
\end{align}
Here the special term 
$\left\langle x_{1}^{[N]}\right| V_{1}^{[N]\dagger }Z^{\otimes N}V_{1}^{[N]}\left| x_{1}^{[N]}\right\rangle $
is given by the following sum of squares, 
\begin{align} \label{eq:special_term}
&\left\langle x_{1}^{[N]}\right| V_{1}^{[N]\dagger }Z^{\otimes N}V_{1}^{[N]}\left| x_{1}^{[N]}\right\rangle  \nonumber \\
&= \sum_{i_{1}<j_{1}}\ldots \sum_{i_{N}<j_{N}}
\left| \sum_{P_{1}} \ldots \sum_{P_{N}} \mathrm{sgn}(P_{1}) \ldots \mathrm{sgn}(P_{N})
\xi ^{(1)}(P_{1}(i_{1},j_{1}),\ldots ,P_{N}(i_{N},j_{N}))
\eta ^{(1)*}(P_{1}(j_{1},i_{1}),\ldots ,P_{N}(j_{N},i_{N})) \right| ^{2},
\end{align}
where $P_{k}(a_{k},b_{k})$ is $a_{k}$ ($\mathrm{sgn}(P_{k})=1$) or $b_{k}$ ($\mathrm{sgn}(P_{k})=-1$).
Terms other than this are obtained by replacing some (or one) $Z$'s in $Z^{\otimes N}$ by $I$.
For example, when the $p$-th $Z$ is replaced by $I$, the corresponding SOS form is obtained by 
dropping $\mathrm{sgn}(P_{p})$, 
replacing $\sum_{i_{p}<j_{p}}$ by $\sum_{i_{p}}\sum_{j_{p}}$, 
$P_{p}(i_{p},j_{p})$ in $\xi ^{(1)}$ by $i_{p}$, and 
$P_{p}(i_{p},j_{p})$ in $\eta ^{(1)}$ by $j_{p}$ in the right-hand side of Eq.~(\ref{eq:special_form}).
One can prove the above results by induction.
Since all terms in the right-hand side of Eq.~(\ref{eq:sum}) are (strictly) positive for $1/d \leq \alpha \leq 1/2$, 
we conclude that $M_{1,1}^{[N]}>0$.

As for the second positivity condition (B), it is very hard to find a closed form for
\begin{equation*}
\Theta ^{[N]} =
\left\langle x_{1}^{[N]}\right| M_{(1,1)}^{[N]} \left| x_{1}^{[N]} \right\rangle 
 \left\langle x_{2}^{[N]} \right| M_{(2,2)}^{[N]} \left| x_{2}^{[N]} \right\rangle
-\left| \left\langle x_{1}^{[N]} \right| M_{(1,2)}^{[N]} \left| x_{2}^{[N]} \right\rangle \right|^{2}.
\end{equation*}
However, if we assume that all variables are real, we can prove the following by induction,
\begin{align*}
\Theta ^{[N=1]} &=\frac{1}{2}\sum_{i=1}^{d}\sum_{j=1}^{d}
\left[ \sum_{k=1}^{d}\left([ijkk]-[jikk]+[kkij]-[kkji]+[kijk]-[ikkj]\right) \right] ^{2} \\
&+\frac{1}{48} \sum_{i=1}^{d}\sum_{j=1}^{d}\sum_{k=1}^{d}\sum_{l=1}^{d}
\left[(g_{1}-g_{3}+g_{5})^{2}+(g_{1}-g_{4}+g_{6})^{2}+(g_{2}-g_{3}+g_{6})^{2}+(g_{2}-g_{4}+g_{5})^{2} \right],
\end{align*}
where
$[ijkl]$ is the abbreviation of $v_i^{(1)}v_j^{(2)}w_k^{(1)}w_l^{(2)}$,
\[
g_{1}=[ijkl]-[ijlk]-[jikl]+[jilk],
\]
\[
g_{2}=[klij]-[klji]-[lkij]+[lkji],
\]
\[
g_{3}=[ikjl]-[iklj]-[kijl]+[kilj],
\]
\[
g_{4}=[jlik]-[jlki]-[ljik]+[ljki],
\]
\[
g_{5}=[iljk]-[ilkj]-[lijk]+[likj],
\]
and
\[
g_{6}=[jkil]-[jkli]-[kjil]+[kjli].
\]
It seems to be very hard to generalize this result to $N \geq 2$ and complex variables case.
The possibility of non SOS character or nonpositivity of $\Theta ^{[N]}$ cannot be excluded.

\section{Conclusions} \label{Conclusions}

A trial to solve the problem of existence of NPPT-BE has revealed the intriguing relation to the Hilbert's 17th problem.
We have witnessed much progress of commutative algebra and real algebraic geometry concerning the Hilbert's 17th problem 
such as Reznik's theorem recently.
One may expect that some of such concepts and techniques developed so far eventually lead to 
the final proof or disproof of the existence of NPPT-BE.




\begin{thebibliography}{99}
\bibitem{comment_1}
Here, we refer to the distillability as that 
with respect to LOCC maps but not that under PPT operations 
\cite{Rains99,Rains01,CDKL,EVWW} 
or under a set of asymptotic separability preserving operations \cite{BP}.
\bibitem{Superactivation}
In the multipartite setting, 
the superactivation has been found \cite{Superactivation_1} 
leading to the nonadditivity of multiparty quantum channel capacity 
\cite{Superactivation_2}.
\end{thebibliography}
\end{document}